\documentclass[aps,twocolumn,superscriptaddress,footinbib,nobalancelastpage,prx]{revtex4-2}
\usepackage{amsmath,amssymb}
\usepackage[colorlinks=true,citecolor=blue]{hyperref}
\usepackage{graphicx}
\usepackage{soul}
\usepackage{braket}
\usepackage{float}
    
\usepackage[table,xcdraw,dvipsnames]{xcolor}

\newcommand*{\gmfld}{\mathcal{F}}
\newcommand*{\df}{D_\gmfld}
\newcommand*{\zf}{\mathbb{Z}_2}
\newcommand*{\zng}{\mathbb{Z}_3}
\newcommand*{\p}{\mathcal{P}}

\newcommand*{\Ham}{\mathcal{H}}

\newcommand{\be}{\begin{equation}}
\newcommand{\ee}{\end{equation}}

\begin{document}
\title{Persistent non-Gaussian correlations in out-of-equilibrium Rydberg atom arrays}

\begin{abstract}
Gaussian correlations emerge in a large class of many-body quantum systems quenched out of equilibrium, as demonstrated in recent experiments on coupled one-dimensional superfluids [Schweigler \emph{et al.}, Nature Physics {\bf 17}, 559 (2021)]. Here, we present a mechanism by which an initial state of a Rydberg atom array can retain persistent \emph{non-Gaussian} correlations following a global quench. This mechanism is based on an effective kinetic blockade rooted in the ground state symmetry of the system, which prevents thermalizing dynamics under the quench Hamiltonian. We propose how to observe this effect with Rydberg atom experiments and we demonstrate its resilience against several types of experimental errors. These long-lived non-Gaussian states may have practical applications as quantum memories or stable resources for quantum-information protocols due to the protected non-Gaussianity away from equilibrium.
\end{abstract}

\author{Aydin Deger}
\affiliation{Department of Physics and Astronomy, University College London, London WC1E 6BT, United Kingdom}
\affiliation{School of Physics and Astronomy, University of Leeds, Leeds LS2 9JT, United Kingdom}

\author{Aiden Daniel}
\affiliation{School of Physics and Astronomy, University of Leeds, Leeds LS2 9JT, United Kingdom}

\author{Zlatko Papi\'c}
\affiliation{School of Physics and Astronomy, University of Leeds, Leeds LS2 9JT, United Kingdom}

\author{Jiannis K. Pachos}
\affiliation{School of Physics and Astronomy, University of Leeds, Leeds LS2 9JT, United Kingdom}

\date{\today}
\maketitle{}

\section{Introduction}

Free field theories describe the dynamics of fields with no interactions. These are known as Gaussian as their path integral description contains some form of Gaussian function \cite{wick1950, Peskin:1995ev}. In the presence of interactions, however, the system's ground state often develops strong non-Gaussian correlations. While Gaussian states are highly structured and can be understood using a variety of theoretical techniques, it is the non-Gaussian states that often play a key role as resources for universal quantum computation and enhancing the efficiency of a range of quantum information protocols, including quantum  teleporation, communication, sensing, metrology and quantum error correction \cite{Browne2003,Braunstein2005,Menicucci2006,Heersink2006,Niset2009,Gomes:2009aa,Mari2012,Walschaers2017,Su2019,Ra:2020aa,Walschaers2021,hebenstreit:2019aa,Verma2022directdetectionof}.

Recent works have investigated the behavior of quantum states under non-equilibrium dynamics, sparked by the intriguing question: What happens to an interacting state when interactions suddenly vanish, e.g., after a global quench of the system? It has been shown that generic closed systems, governed by quadratic Hamiltonians, swiftly relax to Gaussian states, regardless of their initial condition in contrast to recent results established in finely tuned open quantum system~\cite{gluza2016,carollo2023nongaussian}. This can be viewed as an example of a ``quantum central limit" theorem~\cite{Cramer2008,Cramer_2010}. To explain the emergence of Gaussianity, several mechanisms have been proposed, such as spatial scrambling and canonical transmutation, the latter suggesting that Gaussian components of the initial system act as a Gaussian bath, suppressing non-Gaussianity \cite{gluza_mechanisms_2022, gluza2016}. These mechanisms have been used to describe the decay of non-Gaussianity in recent experiments on $\mathrm{{}^{87}Rb}$ superfluids trapped in a double well potential~\cite{Schweigler:2017aa,Schweigler2021}. 
While these studies have provided crucial insights into the process of relaxation in quantum many-body systems~\cite{Gogolin_2016}, they are restricted to systems with effectively non-interacting degrees of freedom, which do not exhibit ``full'' thermalization but only relax towards a Generalized Gibbs Ensemble~\cite{PolkovnikovRMP}. It is thus important to understand the role of Gaussianity in interacting systems, which can exhibit chaotic dynamics and thermalization. In particular, it is important to understand if and how non-Gaussianity could be protected in such many-body systems when they are taken out of their equilibrium state. 

\begin{figure}[b]
	\centering
	\includegraphics[width=1\columnwidth]{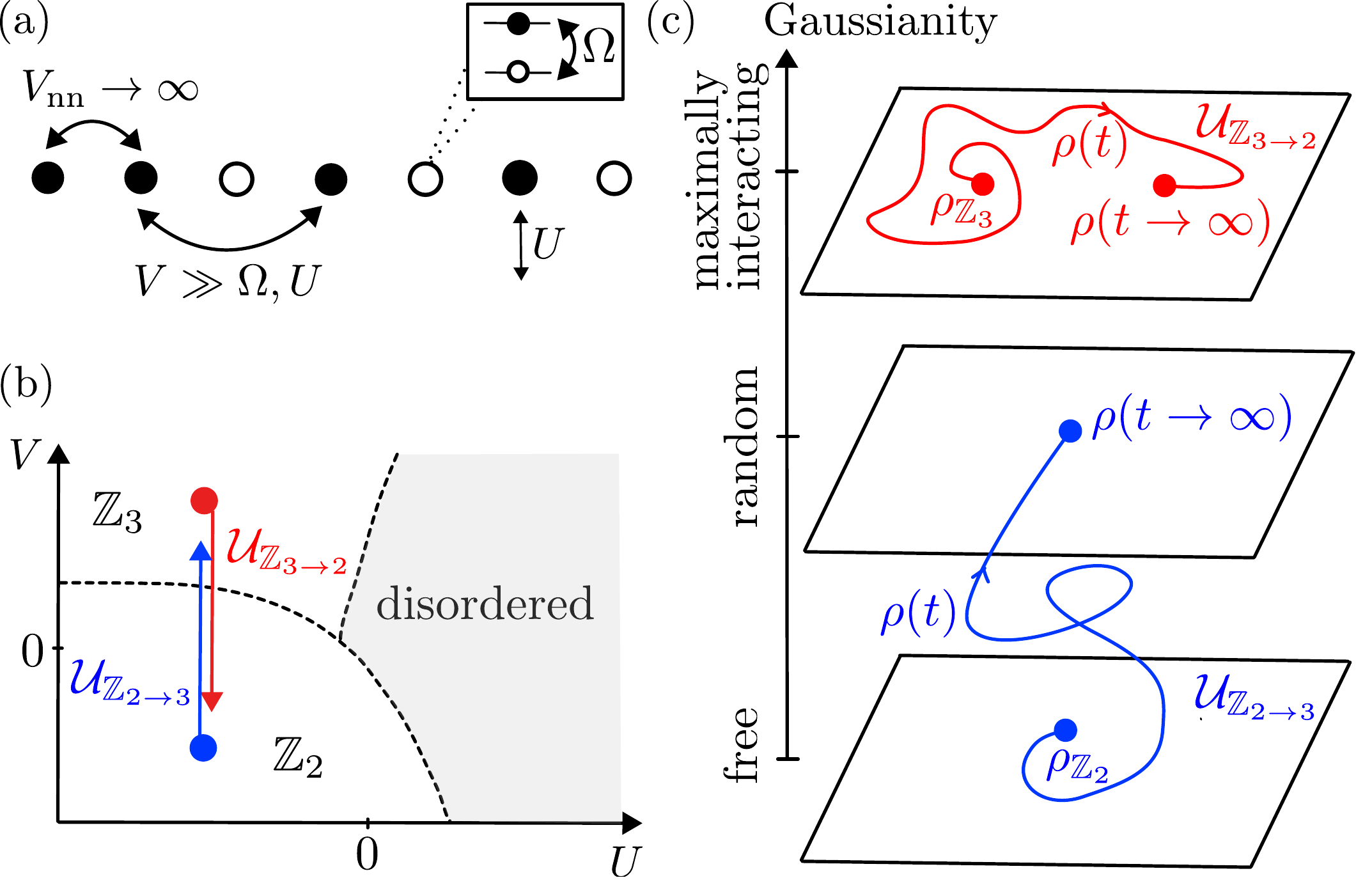}
	\caption{(a) Schematic description of a 1D Rydberg atom model. (b) Sketch of the phase diagram in the $U$-$V$ plane.  We focus on the  $\zf$ and $\zng$ ordered phases and quenches between them: blue arrow $\mathcal{U}_{\mathbb{Z}_{2\rightarrow3}}$ indicates a quench from $\zf$ ordered phase into the $\zng$ phase, while the red arrow represents the reverse quench $\mathcal{U}_{\mathbb{Z}_{3\rightarrow2}}$. (c) The two types of quenches lead to strikingly different dynamical behaviors. During the $\mathcal{U}_{\mathbb{Z}_{2\rightarrow3}}$ quench, the system is initially free and Gaussianity grows until it reaches typical values in a random state. By contrast, during the  $\mathcal{U}_{\mathbb{Z}_{3\rightarrow2}}$ quench, the state remains strongly interacting and it is pinned to a highly non-Gaussian manifold. We will quantify this picture in Secs.~\ref{sec:phasediagram} and \ref{sec:dynamics} using precise Gaussianity measures.
    }
	\label{fig:schematic}
\end{figure}

In this paper, we show that Rydberg atom arrays~\cite{Browaeys2020} provide a versatile experimental platform for realizing and manipulating non-Gaussian correlations far from equilibrium. We show that quenching the atoms  between different ordered phases allows to explore two very different regimes of correlations. On the one hand, our setup allows to observe how non-Gaussian correlations build up as the system undergoes thermalizing dynamics from an initial, nearly free-fermion state. On the other hand, it is possible to ``lock'' the system in a strongly non-Gaussian state, which evades both Gaussification and thermalization at late times. This intriguing non-Gaussian regime is found to be remarkably robust, e.g., even to quenching the system across a quantum phase transition. Our proposal can be readily implemented in Rydberg atom experiments~\cite{Labuhn2016, bernien:2017aa,Keesling:2019aa, omran:2019aa}, which have recently realized the required types of ordered states and protocols for probing correlations in out-of-equilibrium dynamics.

The remainder of this paper is organized as follows. In Sec.~\ref{sec:model} we introduce our model of Rydberg atom arrays and summarize our main results, which are illustrated in Fig.~\ref{fig:schematic}. In Sec.~\ref{sec:phasediagram} we introduce our approach for quantifying Gaussianity, using two complementary diagnostics: the Wick decomposition in terms of local correlations that are experimentally accessible, as well as a more general measure based on reduced density matrix and variational optimization~\cite{Turner:2017aa}. We apply these tools to obtain the Gaussianity phase diagram of a one-dimensional Rydberg atom array. In Sec.~\ref{sec:dynamics} we study dynamics of Gaussianity under global quenches connecting different regions of the phase diagram, whose underlying mechanism is eluciated in Sec.~\ref{sec:mechanisms}. In Sec.~\ref{sec:experiment} we demonstrate the resilience of our results in the face of potential experimental errors, such as local impurity potentials and longer-range interactions. Our conclusions are presented in Sec.~\ref{sec:conclusions}, while Appendixes contain technical details on the choice of operators in Wick's decomposition, finite-size scaling analysis and role of the boundary conditions. 

\section{The model}\label{sec:model}

We consider a one-dimensional (1D) periodic chain containing $N$ Rydberg atoms. Each atom is modelled as a two-level system, where $|0\rangle$ represents an atom in the ground state and $|1\rangle$ is an excited (Rydberg) state. The atomic array is governed by the Hamiltonian~\cite{FendleySachdev}:
\begin{equation}
 \mathcal{H}=\sum_ {i=1}^{N} -\Omega \p_{i-1}\sigma^x_i \p_{i+1} + U n_i + V n_i n_{i+2},
\label{eq:H}
\end{equation}
where $\sigma^x {=} |0\rangle \langle 1 | + |1\rangle \langle 0|$ is the Pauli $x$ operator, $n {=} |1\rangle \langle 1|$ is the local density operator, and $\mathcal{P} {=} |0\rangle \langle 0|$ is the projector on the ground state at a given site. The flipping between the ground and excited states is described by the Rabi frequency $\Omega$, $U$ is the chemical potential (detuning), and $V$ is the next-nearest neighbor interaction between atoms excited to Rydberg states, see Fig.~\ref{fig:schematic}(a). Unless specified otherwise, due to the periodic boundary conditions, we will restrict our calculations to the zero momentum sector. 

The model in Eq.~(\ref{eq:H}) is applicable to the strong Rydberg blockade regime, where the magnitude of the nearest-neighbor van der Waals interaction is much larger than all other couplings of the model.~\cite{Browaeys2020}. The strong Rydberg blockade imposes the kinetic constraint $n_{i} n_{i+1}=0$, which forbids Rydberg excitations on adjacent sites. This constraint is enforced in the Hamiltonian by dressing the $\sigma_i^x$ with projectors $\p = |0\rangle \langle 0|$ on the neighboring atoms. This prevents the Rabi flip term from generating nearest-neighbor excitations, such that states $\cdots 11 \cdots$ with neighboring atoms simultaneously excited are projected out of the Hilbert space. The constraint results in the effective model in Eq.~(\ref{eq:H}), which is also known as the PXP model~\cite{Lesanovsky2012, TurnerQMBS}.

The interplay of $U$ and $V$ terms gives rise to a rich phase diagram sketched in Fig.~\ref{fig:schematic}(b). The phase diagram was mapped out with high precision in the numerical simulations in Refs.~\cite{samajdar_numerical_2018, Rader2019, Chepiga2019} and explored in experiments~\cite{bernien:2017aa, Keesling:2019aa}. Large negative values of the chemical potential $U$ favor excitations on every other site (due to the Rydberg blockade, this is largest density of excitations allowed). On the other hand, positive $V$ value assigns a repulsive potential on the next-nearest neighbors and it favors excitations on every third site. Conversely, large negative $V$ favors excitations on every other site. Thus, the model in Eq.~(\ref{eq:H}) hosts two ordered phases represented by states 
\begin{eqnarray}\label{eq:z2z3}
 \ket{\zf} &=& \frac{1}{\sqrt{2}} (\ket{10101\ldots}+\ket{01010\ldots}), \\
 \ket{\zng} &=& \frac{1}{\sqrt{3}} (\ket{100100\ldots}+\ket{010010\ldots}+\ket{001001\ldots}),   \;\;\;\;\;
\end{eqnarray}
in which Rydberg excitations occupy every second or third site, respectively. Both of these phases are destroyed by sufficiently large positive $U$, which drives the system into a disordered phase, see Fig.~\ref{fig:schematic}(b) \cite{baxter_hard_1980,FendleySachdev,Slagle2021}.

For the subsequent calculations, unless specified otherwise, we set $\Omega=1$ and concentrate on the Gaussianity and entanglement properties of the initial state as we quench the Hamiltonian between $\zf$, $\zng$ ordered phases, Fig.~\ref{fig:schematic}(b). We will show that the choice of the initial state and realization of the quench can have dramatically different influence on the Gaussianity, as illustrated in Fig.~\ref{fig:schematic}(c). For the quench initialized in the $\mathbb{Z}_2$ phase, indicated by $\mathcal{U}_{\mathbb{Z}_{2\rightarrow3}}$ in Fig.~\ref{fig:schematic}(b), the Gaussianity, as precisely defined in Sec.~\ref{sec:phasediagram} below, is initially low because the pre-quench state can be approximately expressed as a free-fermion state. After the quench, the state becomes progressively more correlated, with its Gaussianity approaching that of a random vector in the same Hilbert space at late times. This behavior is consistent with thermalization dynamics. In contrast, the ground state in the $\mathbb{Z}_3$ phase cannot be expressed as a free-fermion state and hence it has high non-Gaussianity. Moreover, following the quench, the state remains strongly interacting, which occurs due to a lack of thermalization in this case. It is important to note that in both cases, the quench Hamiltonian, regardless of the nature of the ground state, is an interacting, non-integrable Hamiltonian -- further contrasting the two regimes. In the following sections, we introduce several metrics of non-Gaussianity and quantitatively support the phase diagram and dynamical behavior sketched in Fig.~\ref{fig:schematic}.

\section{Gaussianity phase diagram}\label{sec:phasediagram}

A conventional approach for quantifying the Gaussianity of quantum states relies on  Wick's theorem~\cite{wick1950}. This theorem allows to reduce the evaluation of $n$-point correlation functions in terms of ``contractions'' (i.e., vacuum expectation values) of pairs of creation and annihilation operators. For any free-fermion system, the Wick's identity for four-point correlators takes the form
\begin{align}
\langle\hat{A} \hat{B} \hat{C} \hat{D} \rangle =\langle\hat{A} \hat{B} \rangle\langle\hat{C} \hat{D} \rangle-\langle\hat{A} \hat{C} \rangle\langle\hat{B} \hat{D} \rangle+\langle\hat{A} \hat{D} \rangle\langle\hat{B} \hat{C} \rangle,
\label{eq:ABCD}
\end{align}
where the expectation value is with respect to the ground state (``vacuum''). One possible definition of Gaussianity $\mathcal{W}$ is the extent to which Eq.~\eqref{eq:ABCD} is violated, i.e., the absolute value of the difference between its left-hand and right-hand side. For Gaussian states, we have $\mathcal{W} =0$. The operators  $\hat A, {\ldots},\hat D$ are understood to be single-site fermionic creation and annihilation operators $\hat f_i, \hat f_j^\dagger$, which obey the anticommutation relation $\{\hat f_i, \hat f_j^\dagger\}=\delta_{ij}$. As our model in Eq.~(\ref{eq:H}) is expressed in terms of spin variables, it will be convenient to work with spin operators rather than fermionic ones, which can be accomplished by applying the Jordan-Wigner transformation. 

In order to distinguish the Gaussianity between $\zf$ and $\zng$ ordered phases, we choose $\hat A= \hat f_1^\dagger$, $\hat B= \hat f_1$, $\hat C=\hat f_2^\dagger$, $\hat D=\hat f_3$, resulting in the following measure of the Wick's decomposition violation:
\begin{align}
\begin{split}
\mathcal{W}(\rho)&=\big|\langle n_1 \sigma_2^{+} \sigma_3^{-} \rangle -\langle n_1 \rangle \langle \sigma_2^{+} \sigma_3^{-} \rangle \\&-
\langle \sigma_1^{+} \sigma_2^{+} \rangle \langle \sigma_1^{-} \sigma_2^z \sigma_3^{-} \rangle+
\langle \sigma_1^{-} \sigma_2^{+} \rangle \langle \sigma_1^{+} \sigma_2^z \sigma_3^{-} \rangle
\big|,
\label{eq:W}
\end{split}
\end{align}
where $\sigma_j^{\pm} \equiv (\sigma_j^x \mp i \sigma_j^y)/2$ are the standard spin raising and lowering operators at site $j$, and $\rho$ denotes the ground state of the system (which could be either a pure state or a density matrix). This particular choice of operators $\hat A, {\ldots}, \hat D$ is justified in Appendix~\ref{appWick}; it will reveal the difference between the $\zf$ phase, where  $\mathcal{W} \approx 0$ across the entire phase, and the $\zng$ phase where the deviation from Wick's decomposition is of order unity, $\mathcal{W} \sim O(1)$.

\begin{figure}
\centering
\includegraphics[width=1\columnwidth]{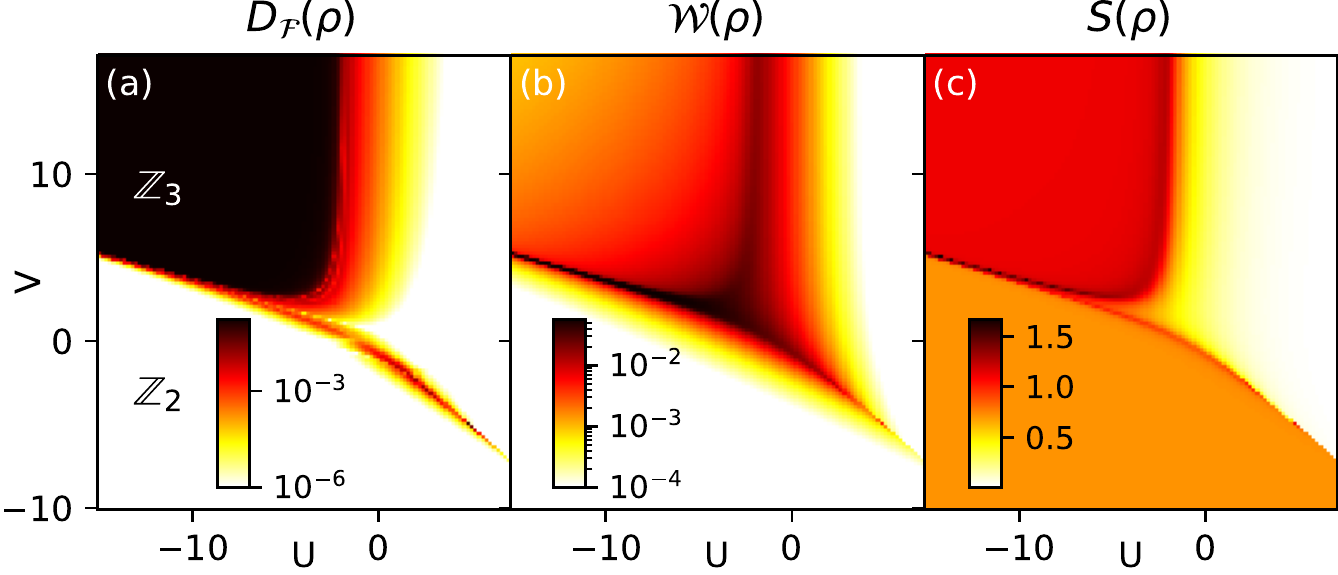}
\caption{Gaussianity phase diagram for the ground state of the model in Eq.~(\ref{eq:H}) as a function of $U$ and $V$. The color scale represents the value of interaction distance $D_\mathcal{F}$ in (a), the Wick's decomposition violation $\mathcal{W}$ in (b), and the entanglement entropy $S$ in (c). All data was obtained by exact diagonalization for $N{=}18$ atoms on a ring with periodic boundary conditions.}
\label{fig:phases}
\end{figure}

The ambiguity in the choice of operators in the Wick decomposition can be eliminated by employing variational optimization techniques to measure the minimum distance between the reduced density matrix of a given state and the set of all density matrices associated with free-fermion models \cite{Turner:2017aa, pachos2018,matos_emergence_2021,Pachos2022quantifying}. This quantity, dubbed the ``interaction distance" $D_\mathcal{F}$~\cite{Turner:2017aa}, allows for a more general characterization of the state's Gaussianity by quantifying its deviation from the \emph{closest} free-fermion model defined in an \emph{arbitrary} basis. Interaction distance is a property of the reduced density matrix $\rho$ describing the subsystem $A$ of a bipartite system $A \cup B$. For the total system in a pure state $|\psi\rangle$, the reduced density matrix $\rho = \text{tr}_B \ket{\psi} \!\bra{\psi}$ is obtained by tracing out the subsystem $B$, and its eigenvalues $\rho_k$ define the so-called ``entanglement spectrum", $\mathcal{E}_k {=} -\ln \rho_k$~\cite{li2008}. Using the entanglement spectrum, the interaction distance is defined as
\begin{equation}
\label{eq:df}
D_{\cal F} (\rho) = \min_{\{ \epsilon \}}\frac{1}{2} \sum_k \left| e^{ - \mathcal{E}_k} - e^{-\mathcal{E}_k^f(\epsilon)} \right|,
\end{equation}
where $\mathcal{E}_k^f(\epsilon)=\sum_l n_l^{(k)} \epsilon_l$ is the entanglement spectrum of a free-fermion system, given in terms of single-fermion modes $\epsilon_l$  and their occupations $n_l^{(k)} \in \{0,1\}$~\cite{Peschel:2003aa}. The sum runs over the many-body entanglement spectrum. The minimization is over the single-particle energies $\{ \epsilon \}$, whose number typically scales linearly with the number of atoms $N$. 

It is worth noting that the entanglement spectrum is naturally dependent on the choice of bipartition. In this work, we consider a bipartition of the system into two equal parts. Our results, however, are not sensitive to the particular choice of partition, as long as both subsystems are of comparable sizes. If one subsystem is much smaller than the other, the number of Schmidt coefficients will be significantly reduced and we expect non-universal behavior of $D_\mathcal{F}$. Interestingly, the interaction distance can also be probed with respect to the eigenspectrum of the system -- thus probing the thermal properties of a given Hamiltonian. This analysis can be done and reveals that in both ordered regimes, the Hamiltonian is interacting. Generally, however, this is more cumbersome to perform with increasing system size due to the exponential scaling and does not reveal the distinct differences in the ground state between the two regimes.

Intuitively, $D_\mathcal{F}$ represents the minimum distance between the reduced density matrix of a given quantum state $\rho$ and the density matrix of the closest free-fermion model defined on the subsystem $A$. Crucially, the free-fermion model is defined up to an arbitrary unitary transformation on $A$, which makes $D_\mathcal{F}$ basis independent. 
This allows to quantify the Gaussianity of a quantum state without the need to search for suitable operators in $\mathcal{W}$ as done in Eq.~(\ref{eq:W}). However, $\mathcal{W}$ has the advantage over $D_\mathcal{F}$  in that it is expressed in terms of local correlations that are amenable to experimental measurements. Thus, Eq.~\eqref{eq:W} provides a more practical way of detecting non-Gaussianity in the lab.

The Gaussianity phase diagram for the ground state of the Hamiltonian in Eq.~\eqref{eq:H} is presented in Fig.~\ref{fig:phases} for a range of $U$ and $V$ values. The phase diagram was obtained using both the interaction distance $D_\mathcal{F}$ and the Wick's theorem violation $\mathcal{W}$, shown in panels (a)-(b). In panel (c) we also show the von Neumann entanglement entropy, $S(\rho) = -\sum_k \rho_k\ln \rho_k$. All the quantities were computed for the reduced density matrix corresponding to the subsystem $A$ being one half of the chain.  Fig.~\ref{fig:phases} reveals excellent qualitative agreement between all three metrics, in particular between interaction distance and Wick's decomposition. The phase boundaries are in good agreement with Refs.~\cite{samajdar_numerical_2018, Chepiga2019}, suggesting weak finite-size effects.  For large and negative chemical potential $U$, there are two competing ordered phases, $\zf$ and $\zng$. In particular, for large and positive values of $V$, the ground state is the $\zng$ ordered state. The quantum phase transition from $\zng$ to $\zf$ ordered state occurs at around $|V|\sim -U/3$. In between these two ordered phases, we expect a narrow intermediate commensurate phase~\cite{FendleySachdev,samajdar_numerical_2018,Keesling:2019aa}, which is difficult to resolve in small systems used in Fig.~\ref{fig:phases}, but this phase will be unimportant for our discussion. 

Figures~\ref{fig:phases}(a)-(b) reveal a stark contrast between the two ordered phases in terms of the Gaussian nature of their ground-state correlations: while the $\zf$ ground state is approximately a non-interacting Gaussian state with both $D_\mathcal{F}$ and $\mathcal{W}$ close to zero, the $\zng$ ground state is nearly maximally interacting, non-Gaussian state. The notion of ``maximally interacting" can be made precise by noting that $\df$, as a trace distance between density matrices, has an upper bound, which has been conjectured to be $\df^{\max}=3-2\sqrt{2}$~\cite{kon2018}. In the $\zng$ phase in Fig.~\ref{fig:phases}, $D_\mathcal{F}$ attains values very close to this upper bound, strongly suggesting that it is not possible to express the $\zng$  ground state as a Gaussian state of free-fermionic modes.  Finally, we note that the entanglement entropy in Fig.~\ref{fig:phases}(c) also captures some features of the phase diagram, but it does not sharply distinguish between the $\zf$ and $\zng$ phases. Thus, the interaction distance and local Wick decomposition are essential to gain a complete understanding of non-Gaussianity, both in equilibrium as well as out-of-equilibrium, as we show next.

\begin{figure}[t]
\centering
\includegraphics[width=.95\columnwidth]{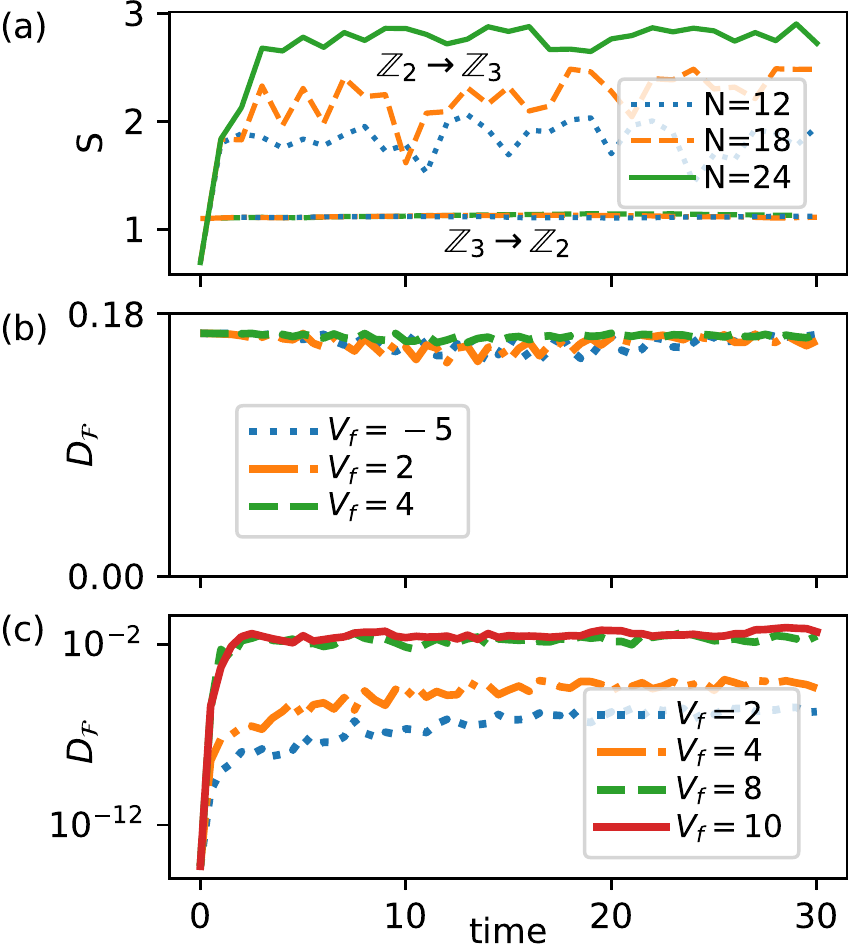}
\caption{Temporal behavior of entanglement and Gaussianity for quenches $\mathcal{U}_{\mathbb{Z}_{2\rightarrow3}}$ and $\mathcal{U}_{\mathbb{Z}_{3\rightarrow2}}$, previously indicated in Fig.~\ref{fig:schematic}(b). The chemical potential is held fixed at $U{=}{-}15$. (a) Growth of entanglement entropy for different system sizes. The top three lines represent $\mathcal{U}_{\mathbb{Z}_{2\rightarrow3}}$ (specifically, $V_\mathrm{i}{=}{-}5 \to V_\mathrm{f}{=}8$), while the bottom three (overlapping) lines are for the reverse $\mathcal{U}_{\mathbb{Z}_{3\rightarrow2}}$  quench. For the $\mathcal{U}_{\mathbb{Z}_{2\rightarrow3}}$ quench, the saturation entropy obeys the volume law scaling with system size, indicating thermalization. By contrast, $\mathcal{U}_{\mathbb{Z}_{3\rightarrow2}}$ quench leads to strongly non-thermalizing dynamics, as evidenced by a complete lack of entropy growth.
(b)-(c) Temporal behavior of Gaussianity measured by interaction distance.  In (b), we quench from the $\zf$  ground state ($V_\mathrm{i}{=}{-}5$) to a range of $V_\mathrm{f}$ values spanning both $\zng$ and $\zf$ phases.  The top plateau value corresponds to the interaction distance of a random state, $\df^\mathrm{random} \approx 0.03$~\cite{kon2018}, consistent with thermalization observed for $\mathcal{U}_{\mathbb{Z}_{2\rightarrow3}}$ in (a). (c) Similar to (b) but for $\zng$ initial state ($V_\mathrm{i}{=}8$). The persistent large value of $D_\mathcal{F}$ is consistent with an absence of entanglement spreading for $\mathcal{U}_{\mathbb{Z}_{3\rightarrow2}}$ quench in (a).
Data in panels (b)-(c) is for system size $N{=}18$. }
\label{fig:rawdfS}
\end{figure}

\section{Persistent non-Gaussian correlations under quench}\label{sec:dynamics}

Previously, we have seen that the two competing ordered phases, $\zf$ and $\zng$, are the extreme points on the Gaussianity spectrum: while the $\zf$ ground state represents a nearly-free fermion state, the $\zng$ state is maximally interacting. It is natural to inquire about the temporal evolution of Gaussianity following a sudden quench between these phases. According to the standard scenario of thermalization in a closed system~\cite{Ueda2020},  under quench dynamics, particularly across a quantum phase transition, the system should lose memory of its initial state and equilibrate towards a maximally entangled state. To test this expectation, we study the spreading of entanglement and non-Gaussian correlations when the system is prepared in the ground state of the Hamiltonian \eqref{eq:H} for some value $V\equiv V_\mathrm{i}$. We then quench the Hamiltonian to some different value of $V\equiv V_\mathrm{f} \neq V_\mathrm{i}$. By varying $V_\mathrm{i}$ and $V_\mathrm{f}$ we can access different ordered states and post-quench Hamiltonians. For simplicity, we keep $U$ the same in the initial and post-quench Hamiltonian and postpone the discussion of its role to Sec.~\ref{sec:experiment}.

Figure~\ref{fig:rawdfS}(a) contrasts the growth of entanglement entropy for the $\mathcal{U}_{\mathbb{Z}_{2\rightarrow3}}$  quench vs. $\mathcal{U}_{\mathbb{Z}_{3\rightarrow2}}$ quench. In the first case, the system exhibits thermalization, as confirmed by the fast growth of entropy towards its saturation value when it reaches the thermal state. 
A key indication of thermalization is the volume-law scaling behavior of the saturation value of entanglement entropy, $S_\infty \propto N$, consistent with Fig.~\ref{fig:rawdfS}(a). 
In contrast, quenches from the $\zng$ state lead to non-thermalizing dynamics, as seen in the strongly suppressed growth of entropy in Fig.~\ref{fig:rawdfS}(a). 

In Fig.~\ref{fig:rawdfS}(b) we illustrate how Gaussianity changes in time when we prepare the system in an approximately Gaussian $\zf$ state at $t=0$. In particular, when the post-quench Hamiltonian is in the $\zng$ phase (e.g., $V_\mathrm{f}=8$), the deviation from Gaussianity sharply increases from zero and quickly reaches the saturation value of $\df^\infty \approx 0.03$, which coincides with interaction distance of a random vector~\cite{kon2018}. This is consistent with thermalizing dynamics at infinite temperature in Fig.~\ref{fig:rawdfS}(a), where the state at late times becomes similar to a random vector. Note that this scenario is very different from Ref.~\cite{Schweigler2021}, where the initial state was chosen to be non-Gaussian, but the Hamiltonian itself is quadratic and induces the development of Gaussian correlations over time.

Conversely, for the non-Gaussian $\zng$ initial state in Fig.~\ref{fig:rawdfS}(c), we see that the previous scenario does not hold.  In this case, there is persistent non-Gaussianity after the quench, with no sign of decay of the correlations due to interactions.  Consequently, the time-evolved state remains highly interacting over the course of quantum dynamics. We note that these results hold for larger system sizes via finite-size scaling and with open boundary conditions -- see Appendix~\ref{appOBC}.

\begin{figure}[t]
\centering
\includegraphics[width=\columnwidth]{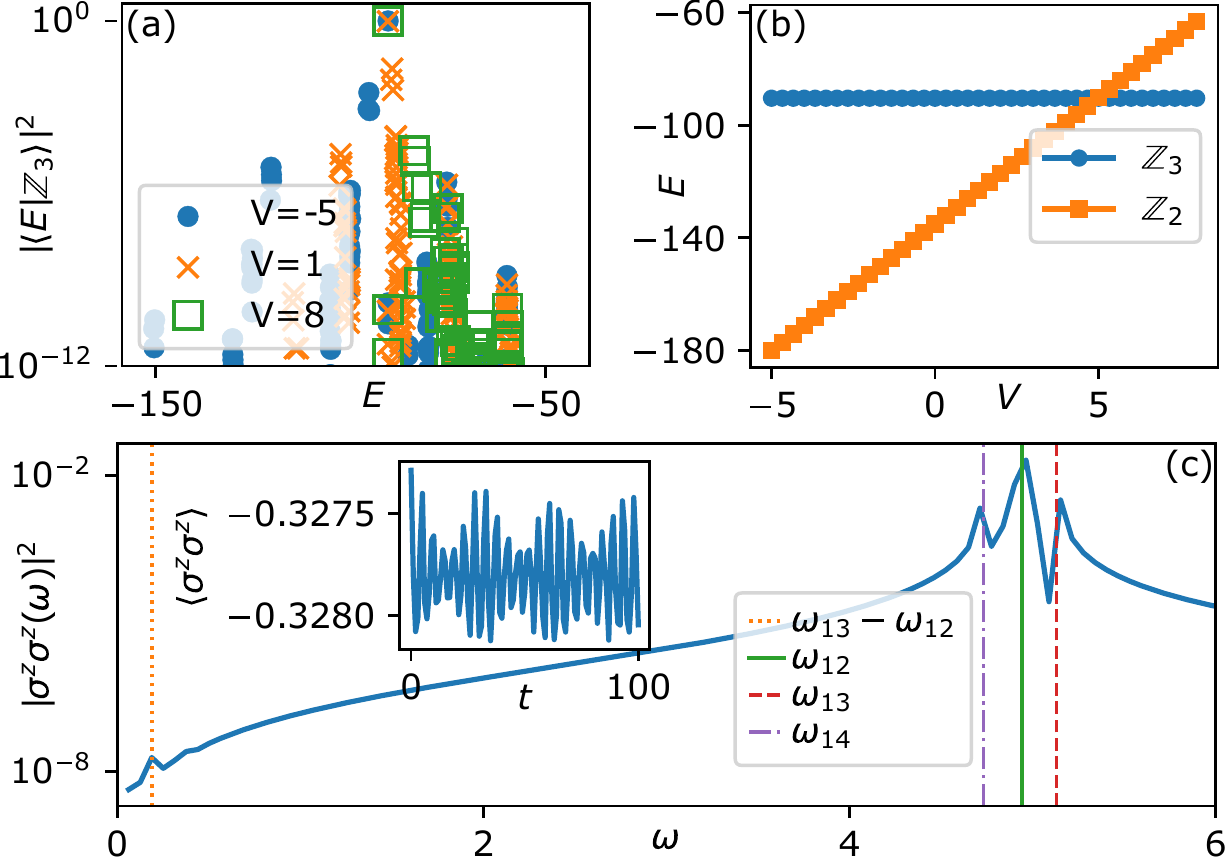}
\caption{The nature of the non-Gaussian quench $\mathcal{U}_{\mathbb{Z}_{3\rightarrow2}}$.
(a) Overlap between the initial $\zng$ state (as defined in Eq.\eqref{eq:z2z3}) and energy eigenstates $\ket{E}$ of the Hamiltonian in Eq.~(\ref{eq:H}), plotted as a function of energy. Data is for system size $N{=}18$ and $U{=}-15$, for three $V$ values given in the legend.  In all the cases, the $\zng$ state has high support (overlap $\approx1$) on a single eigenstate at roughly the same energy.
(b) Energy expectation value, $E=\left \langle \psi|\mathcal{H}|\psi \right \rangle$, for product states $\ket{\psi}=\ket{\zng}$ and $\ket{\psi}=\ket{\zf}$, plotted as a function of $V$ with fixed $U{=}-15$. We see that $\ket{\zng}$ remains at constant energy for any $V$, while $\ket{\zf}$ scales linearly. (c) The power spectrum of the correlation function $\langle \sigma^z_i \sigma^z_{i+1} \rangle$ evaluated in the time evolved state (raw data shown in the inset). The vertical lines in the plot represent the energy gaps $\omega_{ij}$ between the states with largest overlap in (a). The gaps align precisely with the peaks in the power spectrum.}
\label{fig:overlap}
\end{figure}

We now analyze the $\mathcal{U}_{\mathbb{Z}_{3\rightarrow2}}$ quench in more depth. In terms of spectral properties, we find that there is a single energy eigenstate $\ket{E_{1}}$ of the Hamiltonian that has very high overlap with the $\zng$ state, $|\braket{E_1|\zng}|^2\approx 1$, see Fig.~\ref{fig:overlap}(a). The energy of this eigenstate also exactly matches that of the ground state energy of the initial Hamiltonian pre-quench. In fact, the energy of the state is independent of deforming $V$ in the quench Hamiltonian as shown in Fig.~\ref{fig:overlap}(b). This implies that the $\zng$ state is effectively close to being an eigenstate of the Hamiltonian. This behavior is somewhat reminiscent of quantum many-body scarring~\cite{Serbyn2021, MoudgalyaReview, ChandranReview}, with the exception that here the high overlap exists only for a single eigenstate. Furthermore, numerically we find that the special eigenstate $\ket{E_{1}}$ has lower entanglement entropy than the majority of the spectrum, whereas its interaction distance attains a nearly maximum value. By also plotting the overlap between the Z3 state and the eigenstates of Hamiltonians with $V=8$ and $V=1$, we see the single eigenstate remains dominant at constant energy, regardless of the value of $V$, simply transitioning from being an initial ground state to a mid-spectrum state.

Previously, in Fig.~\ref{fig:rawdfS}(c), we saw that non-Gaussianity remains robust for the quench $\mathcal{U}_{\mathbb{Z}_{3\rightarrow2}}$. In order to experimentally access this behavior, one can study temporal behavior of local correlation functions, as frequently done in modern ultracold atom experiments~\cite{BlochRMP}. For example, the correlation function $\langle \sigma_i^z \sigma_{i+1}^z\rangle$, computed in Fig~\ref{fig:overlap}(c), reveals persistent oscillations. The characteristic frequencies of these oscillations correspond to the energy differences of the eigenstates with dominant overlap with the initial state of the system. This can be characterized more precisely by the power spectrum~\cite{Kormos:2017aa,Chen2021}, computed in Fig.~\ref{fig:overlap}(c). The dominant frequencies are given by $\omega_{1j}=|E_1-E_j|$ (and their differences), where $\ket{E_j}$, $j={2,3,\ldots}$, denote eigenstates with subleading overlaps with the $\zng$ state.  Similar oscillations and frequencies can be observed in the quantity $\mathcal{W}$  defined in Eq.~\eqref{eq:W}, other two-point local correlations, and even in the entanglement entropy.

A simple heuristic argument that gives an approximate value of the oscillation frequency in the limit $U,V \gg \Omega$ can be stated as follows. For the Hamiltonian with $U{=}{-}15$ and $V{=}{-}5$, the ground state is approximately $\zf$ product state with energy $E^{\zf}_{\rm GS} \approx (U+V)N/2$. The energy of the $\zng$ state is $E^{\zng} \approx UN/3$. For general values of $U$ and $V$ there are no other states with the same energy as $\zng$. For special ratios of $U/V$, a resonance may occur and other states could have the same energy as $\zng$; we can prevent this by assuming $U$ and $V$ to be irrational numbers. The oscillations seen in Fig.~\ref{fig:overlap}(c) are between $\ket{\zng}$ and states where one of the excitations is moved by a single unit, i.e., the states $\ket{101000100100\ldots}$, $\ket{100101000100\ldots}$ etc., which all have a single $101$ pattern. The energy of these states is $UN/3 + V$, so they are lower in energy by approximately $-|V|$ compared to $\zng$. This predicts that the oscillation frequency is set by $V$, i.e., the energy difference between $\zng$ states and these states containing $\ket{\dots 101 \ldots}$ in the chain. Thus the energy differences between second and first energy levels are determined by $|V|$, i.e., $\omega_{12}=|E_1-E_2|\approx|V|$, as can be seen in the power spectrum in Fig.~\ref{fig:overlap}(c).

\section{Origin of persistent non-Gaussian correlations} \label{sec:mechanisms}

The origin of robust non-Gaussianity associated with the $\zng$ state can be more readily understood by considering the evolution in eigenspace overlap for different $V$ presented in Fig.~\ref{fig:overlap}(a). We consider the difference between the initial Hamiltonian, $\mathcal{H}_i$, and post-quench Hamiltonian, $\mathcal{H}_f$. Recalling Fig.~\ref{fig:schematic}(b), we restricted to the case where the quench only changes the value of $V$. Therefore $\Ham_f=\Ham_i + \Delta V \Ham^{nn}$ where $\Ham^{nn}=\sum_i n_in_{i+2}$. Now consider a quench from the $\zng$ state in its respective regime such that $\Ham_i\ket{\zng}=E_0\ket{\zng}$ with ground state energy $E_0$. Then, quenching yields $\Ham_f\ket{\zng}=\Ham_i\ket{\zng} + \Delta V\Ham^{nn}\ket{\zng}$. Note that in the case of $\ket{\zng}$, there is only occupancy of every third site, therefore $\Ham^{nn}\ket{\zng}=0$, irrespective of $\Delta V$. Thus, 
\begin{eqnarray}
\Ham_f\ket{\zng}=\Ham_i\ket{\zng} + \Delta V\Ham^{nn}\ket{\zng} = \Ham_i\ket{\zng} = E_0\ket{\zng}. \;\;
\end{eqnarray}
Hence, upon deforming $\Delta V$, $\ket{\zng}$ remains an eigenstate of $\Ham_f$ with the same energy. This may not necessarily still be the ground state and instead may be shifted up in the energy spectrum for sufficiently large $\Delta V$. This means that upon quenching, the initial state remains the same over long periods of time due to its proximity to an eigenstate. As interaction distance is defined only with respect to a given state $\rho$, it is clear why it does not significantly change over time, despite quenching the system across criticality. This interpretation is supported by the high overlap with a single eigenstate of the quench Hamiltonian in Fig.~\ref{fig:overlap}(a). 

A similar argument can be made for why the quench from the initial $\zf$ state with $\Ham_f$ in the $\zng$ phase leads to scrambling and thermalization dynamics. As $\zf$ has an occupancy on every 2 lattice sites, it ``feels" the deformation of $\Delta V$: $\Ham^{nn}\ket{\zf}=\frac{N}{2}\ket{\zf}$. Considering the form of $\Ham_i$ such that the initial state is $\ket{\zf}$, the terms that result in this being approximately the ground state are $\sum^N_{i=1}-|U|n_i-|V|n_in_{i+2}$ with $U,V\gg\Omega$. When then quenching with $\Ham_f$ in the $\zng$ regime, suitable tuning of positive $\Delta V$ then means that given $\Ham_f$, $\ket{\zf}$ is no longer an eigenstate due to the competing factors of $U,V$ and $\Delta V$. More concretely, the final Hamiltonian when acting on $\ket{\mathbb{Z}_2}$ has a term proportional to $-(|U|+|V|) + \Delta V$. These competing negative and positive terms mean that, overall, $-(|U|+|V|) + \Delta V$ may not be much greater than $\Omega$ and $\ket{\zf}$ may no longer be approximately an eigenstate like before. This results in the possible scrambling of the initial state into a non-Gaussian state over time.

The further substantiate the previous argument, the underlying mechanism for persistent non-Gaussian correlations can be inferred by considering an effective quench Hamiltonian with five-body interactions. As we are interested in the quench dynamics going from $\zng$ into the $\zf$ ordered phase, we define the effective Hamiltonian in a regime where $U$ is negative and large,  favoring particle creation on all sites. Similarly, we require $V$ to be large and negative as well. Under these conditions, the quench Hamiltonian is given by
\be
\mathcal{H}_{q}=- \left[\sum_ {i=1}^{N} \p_{i-1}\sigma^x_i \p_{i+1}+ |U| n_i + |V| n_i n_{i+2}\right]
\label{eq:Hq}
\ee
with $|U|{\gg}1$, $|V|{\gg}1$ and we still have $n_i n_{i+1}{=}0$. Following a similar procedure as in Ref.~\cite{igor2011}, we move the quench Hamiltonian into an interaction picture with respect to the next-nearest neighbor term by applying the transformation $W^\dagger \mathcal{H}_{q} W$, where $W=\exp[-i t |V| \sum_i n_{i} n_{i+2}]$. Ignoring the rapidly oscillating phases for $|V|\gg1$, we reach an effective Hamiltonian
\be
\mathcal{H}_{q}^{\rm eff} = - \left[ \sum_ {i=1}^{N} \p_{i-2}\p_{i-1} \sigma^x_i \p_{i+1}\p_{i+2} + |U|n_i \right].
\label{eq:Hqeff}
\ee
In the largest fully connected sector, the presence of Rydberg excitations on the nearest and next-nearest neighboring sites is prohibited. The effective Hamiltonian corresponds to the PPXPP model with a chemical potential. The $\zf$ state still exists as the overall ground state but instead within a small disconnected sector due to the new blockade condition. Meanwhile, due to a large negative $U$ in the effective Hamiltonian, the ground state of the largest sector where the blockade remains respected is $\zng$. Thus, the quench Hamiltonian does not induce delocalizing dynamics when the system is initialized in the $\zng$ state and such states are protected against both Gaussification and thermalization. We can therefore conclude that the persistent non-Gaussianity of the $\zng$ initial state equivalently arises from the effective blockade mechanism up to the next-nearest neighbor excitations in the interaction picture. Agreement in dynamics was also tested and found numerically between the exact Hamiltonian and effective Hamiltonian. This further supports the notion that the initial state remains approximately an eigenstate of the quench Hamiltonian. 

\section{Experimental implications} \label{sec:experiment}

With the possible quantum information applications, it is important to test the robustness of the non-Gaussification against external perturbations. This is particularly important because our results rely on the quantum superpositions of states with degenerate energies in the $\zng$ and $\zf$ phases. External perturbations may result in the superposition collapsing into an energetically favorable product state, thus removing any non-Gaussian correlations. Here we focus on three types of effects that are relevant for experimental implementations: (i) the stability against a single site magnetic field or impurity $\varepsilon  n_i$ with magnitude $\varepsilon$; (ii) the effect of changing the chemical potential $U$ during the quench; (iii) the effect of long-range van der Waals interactions that are present in real systems but were neglected in Eq.~(\ref{eq:H}). 

\begin{figure}
\centering
\includegraphics[width=1\columnwidth]{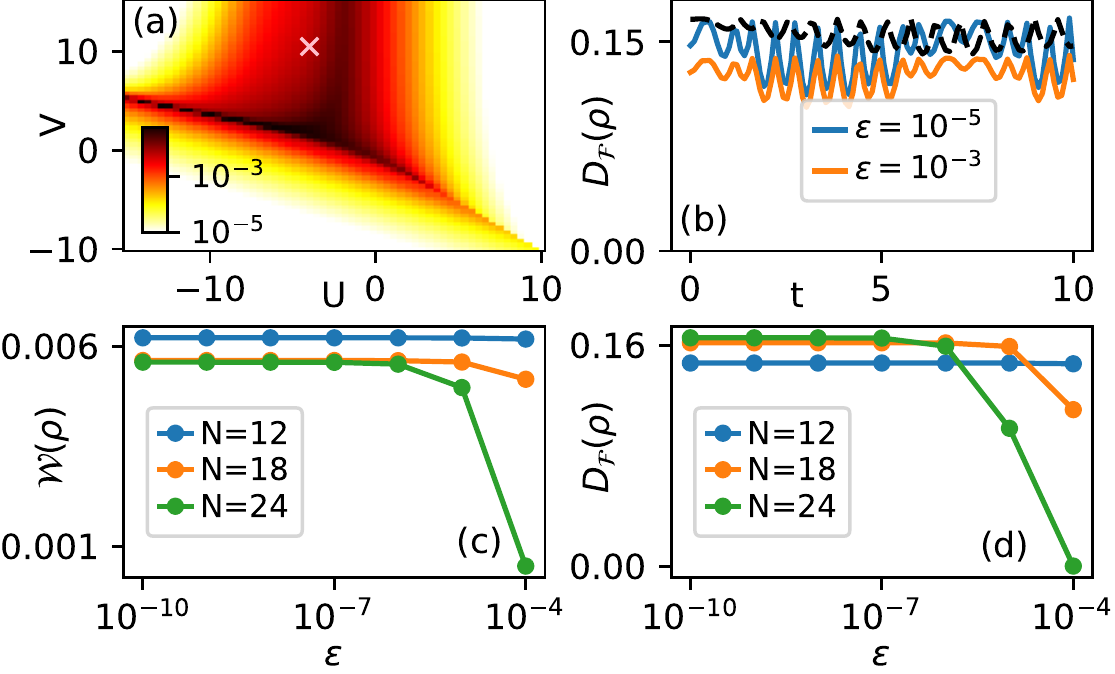}
\caption{Resilience of non-Gaussianity against a local impurity, $\varepsilon  n_4$, added to Eq.~\eqref{eq:H}. (a) The phase diagram obtained using local Wick's decomposition, $\mathcal{W}(\rho)$, with impurity strength  $\varepsilon = 10^{-4}$ at system size $N=12$. The cross marks the point $(U,V){=}(-4,10.5)$ studied in (b)-(d). (b) The time evolution of the interaction distance $D_\mathcal{F}$, when quenching the initial ground state at $(U,V){=}(-4,10.5)$ in the $\zng$ phase with the Hamiltonian parameters $(U,V){=}(-4,-6)$ in the $\zf$ phase. Both the initial and quench Hamiltonian contain impurity $\varepsilon $. Data is for system size $N=12$ with the impurity potentials shown in the legend. Black dashed line is with no error ($\varepsilon=0$) but instead taking the initial ground state at $(U,V){=}(-15,8)$ and quenching at $(U',V'){=}(-10,-5)$, which demonstrates persistent non-Gaussinity even with a change in $U$. (c)-(d) Wick's decomposition and interaction distance, respectively, of the ground state at $(U,V){=}(-4,10.5)$, as a function of impurity strength for several system sizes. The results were computed with exact diagonalization without resolving translation symmetry.}
\label{fig:z3error}
\end{figure}

Fig.~\ref{fig:z3error} shows the results obtained when we add the impurity term $\varepsilon n_i$ to the Hamiltonian in Eq.~\eqref{eq:H} on site $i=4$. We choose this site along the chain as it is found to have the most substantial effect on the results providing a qualitative lower-bound in robustness. Despite the presence of an impurity, we see that qualitative features of the phase diagram remain preserved with an impurity strength $\varepsilon{=}10^{-4}$, see Fig.~\ref{fig:z3error}(a). This is a magnitude of error much larger than the detuning resolution of current quantum technology~\cite{wurtz2023aquila}. Furthermore, perturbations are generally characterised by their proportionality to the ground state gap. We find this order of magnitude to be comparable to the energy gap of the system (which decreases with $N$). This demonstrates that the non-Gaussian characteristics are protected nearly up to the same order as the energy gap in the system. It is natural that any larger magnitude of error would disrupt this as one would no longer be probing ground-state physics. Taking a single point in this diagram, marked by the cross, we find the ground state still possesses high overlap with the superposition state $\zng$. Consequently, the non-Gaussian correlations persist when quenching in the $\zf$ phase (with error still present in the quench Hamiltonian), as seen in Fig.~\ref{fig:z3error}(b). For this point, the non-Gaussianity remains robust for impurity strengths up to $\varepsilon {\sim} 10^{-3}$. Furthermore, in Figs.~\ref{fig:z3error}(c)-(d) we test the robustness of $D_\mathcal{F}$ and $\mathcal{W}$ for this point with varying impurity strength and system size. The non-Gaussianity is seen to be more pronounced in smaller system sizes. 

While our presented analysis assumed that only $V$ is changed during the quench, we have numerically verified that the non-Gaussian correlations also remain robust upon simultaneous changes in $U$. This can be understood via the following argument. Consider modulating both $V$ and $U$, then $\Ham_f\ket{\zng}=\Ham_i\ket{\zng} + \Delta U \Ham^{n}\ket{\zng}$ where $\Ham^{n}=\sum_i n_i$. Unlike $\Ham^{nn}$, $\Ham^{n}\ket{\zng}\neq0$. This is instead equivalent to quenching horizontally in the phase diagram in Fig.~\ref{fig:phases}; therefore, if $\Delta U$ is such that one remains in the regime where $\ket{\zng}$ is approximately the ground state, the state remains an eigenstate and the non-Gaussianity remains robust. This is illustrated by the black dashed line in Fig.~\ref{fig:z3error} where in changing $V$ during the quench, we also take $\Delta U=5$. On the other hand, if $\Delta U$ is large enough to transition from the $\zng$ regime, thermalization occurs. This stability against small changes in $U$ makes the non-Gaussianity effect robust against possible experimental imperfections.

\begin{figure}
\centering
\includegraphics[width=\columnwidth]{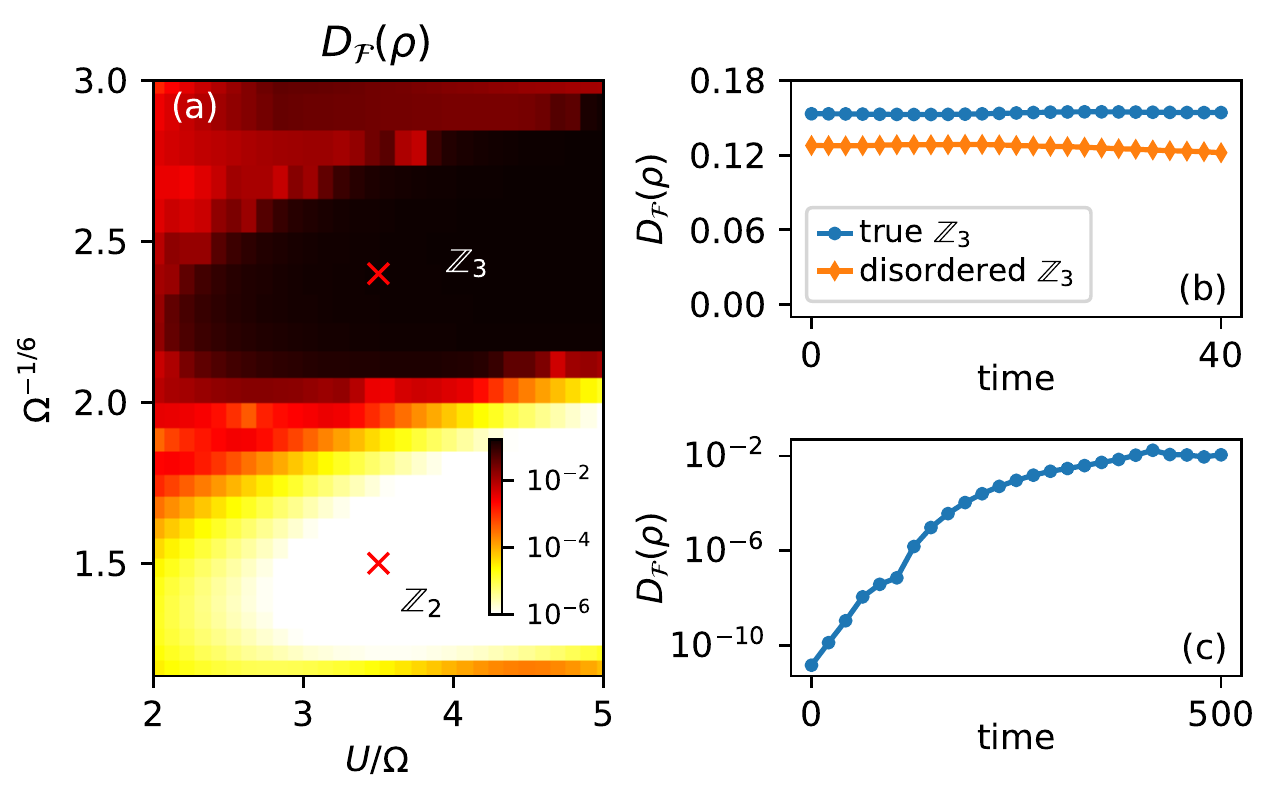}
\caption{(a) Gaussianity phase diagram of the long-range Rydberg model in Eq.~(\ref{eq:rydbergH}) for system size $N{=}12$ and fixed $V{=}1$. (b)-(c) Temporal behavior of $\df$ when quenching from the point indicated by the red cross in the $\zng$ ($\zf$) phase to the other phase. The different behavior of non-Gaussianity for the two types of quenches, seen in Fig.~\ref{fig:rawdfS}, is reproduced. Blue lines indicate using the pure Rydberg Hamiltonian Eq.~\ref{eq:rydbergH}. Meanwhile the orange line introduces further experimental error by using a Hamiltonian with spatial disorder such that $i,j \to i+\delta_i,j+\delta_j$ where $\delta$ is a site dependent and taken randomly in the range $[-0.02,0.02]$. The results are averaged over 100 realisations. Both the pure and disordered Rydberg Hamiltonian show persistent non-Gaussinity.}
\label{fig:rydbergQuench}
\end{figure}

Finally, our idealized model in Eq.~(\ref{eq:H}) neglects the long-range van der Waals forces that are invariably present in real systems of Rydberg atoms~\cite{bernien:2017aa,Keesling:2019aa,omran:2019aa}. Thus, it is important to verify our conclusions still hold in the full model describing the Rydberg atom experiments~\cite{Rader2019}:
\begin{equation}
H = -\frac{\Omega}{2} \sum_{i}^{N} \sigma_{i}^{x} -  U \sum_{i}^{N} n_{i} + V \sum_{i<j} \frac{n_{i}n_{j}}{|i-j|^6}.
\label{eq:rydbergH}
\end{equation}
Note that, in contrast to Eq.~(\ref{eq:H}), here we keep the factor 1/2 in the Rabi term and set $V{=}1$ in order to facilitate comparison with the literature. 
 In Fig.~\ref{fig:rydbergQuench}(a), we first recompute the Gaussianity phase diagram of the long-range model with relevant parameters taken from the experimental papers~\cite{Keesling:2019aa,omran:2019aa}. Similar to the truncated model in Eq.~(\ref{eq:H}), the full model also  realizes both $\zng$ and $\zf$ phases. The phase diagram in Fig.~\ref{fig:rydbergQuench}(a) is in good agreement with that given in Ref.~\cite{Rader2019}.  We then prepare the state in one phase (indicated by a red cross) and perform a quench into the other phase. As illustrated in Figs.~\ref{fig:rydbergQuench}(b)-(c), the results are consistent with those of the UV model, where the $\zng$ state preserves its non-Gaussian correlations. For the $\zf$ initial state, the thermalization time scale is longer than in Fig.~\ref{fig:rawdfS} due to the smallness of the energy gap in the chosen units for the Hamiltonian (\ref{eq:rydbergH}). Taking this a step further, we introduce experimental error into our calculation. Ref.~\cite{wurtz2023aquila} states that there are approximate errors of $\approx0.1\mathrm{\mu m}$ in the spatial position of sites along the Rydberg chain. We can factor this into our simulations by modulating $i,j\to i+\delta_i,j+\delta_j$ in Eq.~\ref{eq:rydbergH} (so, numerically, $\delta$ is randomly sampled from a normal distribution between $\pm0.02$). We find that the results still hold well when taking the initial ground state from a disordered Hamiltonian with only a slight decrease in $D_\mathcal{F}$ as shown by the orange line in Fig.~\ref{fig:rydbergQuench}(b). More-so, if one assumes the perfect $\zng$ state can still be prepared, we find the perfect results still hold irrelevant of the disordered quench Hamiltonian -- adding a degree of robustness as it demonstrates the error only factors into the initial state preparation.  Overall, the main features of our results are present in the full Rydberg model, suggesting that persistent non-Gaussianity could be observed with the existing experimental technology~\cite{Keesling:2019aa, omran:2019aa}. 

\section{Conclusions} \label{sec:conclusions}

Quantum states can exhibit a Gaussian or non-Gaussian nature, depending on the degree of interaction between the system's constituent parts. In this work, we have investigated the phenomenon of Gaussification, wherein non-Gaussian states undergo transformation into Gaussian states during quench dynamics in quantum many-body systems. We have shown that Rydberg atom arrays provide a versatile platform where this behavior can be probed with available experimental techniques. Perhaps more intriguingly, we have demonstrated that the Rydberg blockade, intrinsic to such systems, gives rise to states with remarkably robust non-Gaussian correlations that persist far from equilibrium, e.g., as the system is quenched across the quantum phase transition. We have elucidated the origin of this behavior by analyzing quenches between $\zf$ and $\zng$ phases, which exhibit either scrambling dynamics or suppression of thermalization due to the effective Rydberg blockade mechanism. We formulated a criterion based on Wick's decomposition that incorporates local correlations, providing a practical method for observing (non-)Gaussianity in experiment. This finding was further corroborated via variational optimization and computing the minimal distance between the reduced density matrix of the ground states belonging to different ordered phases and the set of all free-fermion density matrices defined on the same subsystem. 

Our results highlight the richness of quantum state complexity in systems evolving under constrained dynamics, and they provide three contributions to the broader quantum information framework. Firstly, our findings show the existence of robust non-Gaussian states in Rydberg systems, which are well-known resources for enhancing quantum information protocols. Recent studies have shown that non-Gaussian states act as magic states \cite{hebenstreit:2019aa}, facilitating the construction of a universal gate set. Our findings propose a route towards accessing and utilising these robust states in the commonly explored Rydberg systems, moving closer to the realisation of universal quantum devices, even in the presence of long-range interactions and local impurity potentials.
Secondly, we demonstrate how Rydberg systems naturally generate and manipulate $\mathbb{Z}_3$ and $\mathbb{Z}_2$ states, which have a stark contrast in terms of their Gaussianity, due to the interplay between detuning and interactions. Particularly, we find that the $\mathbb{Z}_3$ state is maximally non-Gaussian while simultaneously robust against thermalization, which could serve as a qutrit basis for quantum memories \cite{Brown2016,goel2023unveiling}. Lastly, our results provide a constructive example that diverges from the typical Gaussification scenario of Refs. \cite{Cramer2008,Gogolin_2016}, thus illustrating the possibility of richer types of dynamical behavior facilitated by the Rydberg blockade.

\begin{acknowledgments}

We acknowledge useful discussions with Kieran Bull and Andrew Hallam. This work was in part supported by EPSRC Grant No. EP/T517860/1.  Statement of compliance with EPSRC
policy framework on research data: This publication is
theoretical work that does not require supporting research data.
Z.P.~acknowledges support by the Leverhulme Trust Research Leadership Award RL-2019-015. This research was supported in part by the National Science Foundation under Grant No. NSF PHY-1748958.

\end{acknowledgments}

\appendix

\section{Motivating the choice of operators in the Wick decomposition \label{appWick}}

We motivate the choice of $\hat A$, $\hat B$, $\hat C$, $\hat D$ in Eq.~\eqref{eq:ABCD}, which yielded results qualitatively similar to the interaction distance across the three phases of the diagram. These operators must be single-site fermionic to provide a valid four-point Wick's decomposition.  In order to express these operators in terms of Pauli matrices, we employ the Jordan-Wigner transformation. This allows us to reinterpret the fermionic creation and annihilation operators at site $j$ as spin-raising and lowering operators, multiplied by the string $\prod_{k<j}\sigma_k^z$. This ensures the fermionic commutation relationship still holds. 

We conducted an exhaustive search to verify Wick's decomposition in the UV model in Eq.~(\ref{eq:H}), specifically where the Wick value in the $\zng$ state exceeds that in $\zf$ state. Interestingly, no two-point Pauli correlations that can indicate Gaussianity in the model were found. This unique feature arises from the Rydberg blockade mechanism, setting it apart from other models, such as the $\zf$ phases in the transverse field Ising model. 

\begin{figure}[tb]
\centering
\includegraphics[width=1\columnwidth]{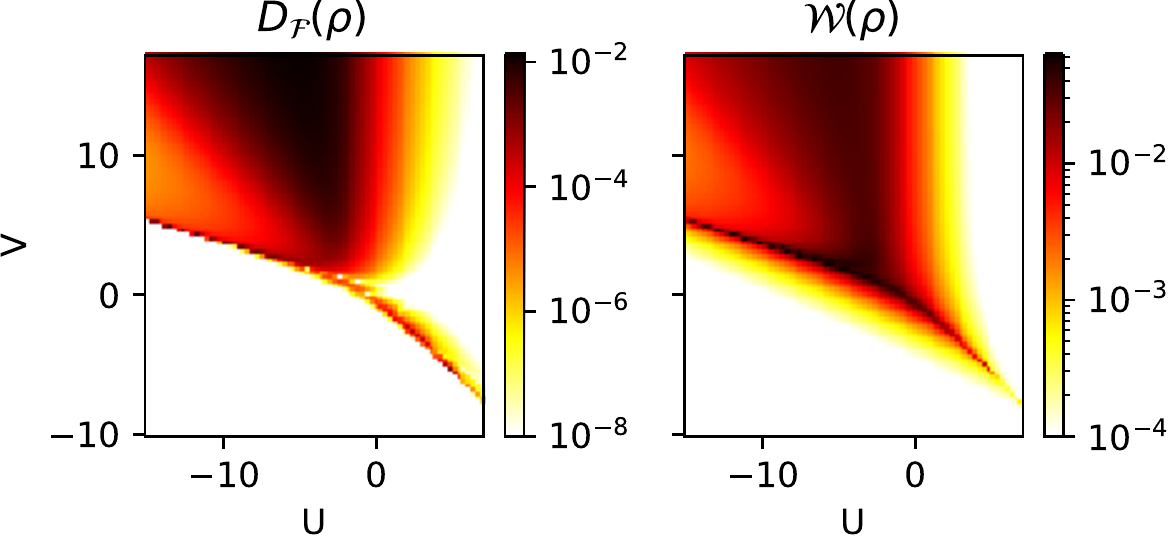}
\caption{Gaussianity phase diagram for open boundary conditions (OBC). The are analogous to Figs.~\ref{fig:phases}(a-b) with interaction distance and Wicks decomposition computed according to Eq.~\eqref{eq:df} and Eq.~\eqref{eq:W}, respectively. System size  $N=15$ and the Wicks decomposition is computed for operators on sites $7,8,9$ in the bulk of the chain. }
\label{fig:suppOBC}
\end{figure}

However, several Wick's decompositions involving 3-point local Pauli decompositions successfully quantify Gaussianity. We discuss a specific choice here: $\hat A=f_1^\dagger$, $\hat B=f_1$, $\hat C=f_2^\dagger$, $\hat D=f_3$. The left-hand side of Eq.~\eqref{eq:ABCD}, $\langle n_1 \sigma_2^{+} \sigma_3^{-}\rangle$, invariably vanishes for the model due to the Rydberg blockade in the case of $\zf$ and due to the action of $\sigma_3^{-}$ on the spin-down state in the $\zng$ regime. Thus the difference in Gaussianity originates from the right-hand side.
To further exemplify and understand this, consider $N=3$ and take both the Greenberger-Horne-Zeilinger (GHZ) state, $\ket{\mathrm{GHZ}}=(\ket{000}+\ket{111})/\sqrt{2}$, (which is a macroscopic superposition like $\ket{\zf}$) and the W state, $\ket{\mathrm{W}}=(\ket{100}+\ket{010}+\ket{001})/\sqrt{3}$ (similar to $\ket{\zng})$. For both states, due to the definition of the spin ladder operators and the form of the superpositions, $\langle n_1 \sigma_2^{+} \sigma_3^{-}\rangle$ and $\langle \sigma_1^{+} \sigma_2^{+} \rangle \langle \sigma_1^{-} \sigma_2^{z} \sigma_3^{-} \rangle$ are zero. This leaves the $\langle n_1 \rangle \langle \sigma_2^{+} \sigma_3^{-} \rangle$ and $\langle \sigma_1^{-} \sigma_2^{+} \rangle \langle \sigma_1^{+} \sigma_2^{z} \sigma_3^{-} \rangle$ components which consist of simple particle hopping terms (with no $n_1$ constraint unlike in $\langle n_1 \sigma_2^{+} \sigma_3^{-}\rangle$). Both of these terms are still found to be 0 in the case of GHZ but non-zero in the W-state due to the fact that we can change from one translated pair to another by simply flipping two spins -- something not possible in the GHZ state due to the macroscopic superposition. This natural distinction between the GHZ and W-state due to simple particle hopping motivates this choice of $\hat A$-$\hat D$ for these states. Recomputing the diagram using purely these two distinguishing hopping terms yields a result that qualitatively matches the interaction distance diagram in Fig.~\ref{fig:phases}.

\section{Open boundary conditions and finite size scaling \label{appOBC}}

We have also examined the Gaussianity phase diagram for open boundary conditions (OBC) and the scaling of the non-Gaussian correlations with system size.  Notably, under OBC, the system no longer retains translational invariance and momentum is no longer a good quantum number. The translation invariance is broken by the boundary terms, which we take to be $\sigma_1^x P_2$ and $P_{N-1}\sigma_N^x$. The dimension of the Hilbert space is also larger than that under PBCs, since states such as $|10101\rangle$ are permissible under OBCs, while the Rydberg blockade precludes these in the PBCs. Moreover, in the OBC context, spontaneous symmetry breaking leads to the absence of exact ground state degeneracy, resulting in the system selecting a nondegenerate ground state. 

\begin{figure}
	\centering
	\includegraphics[width=.5\textwidth]{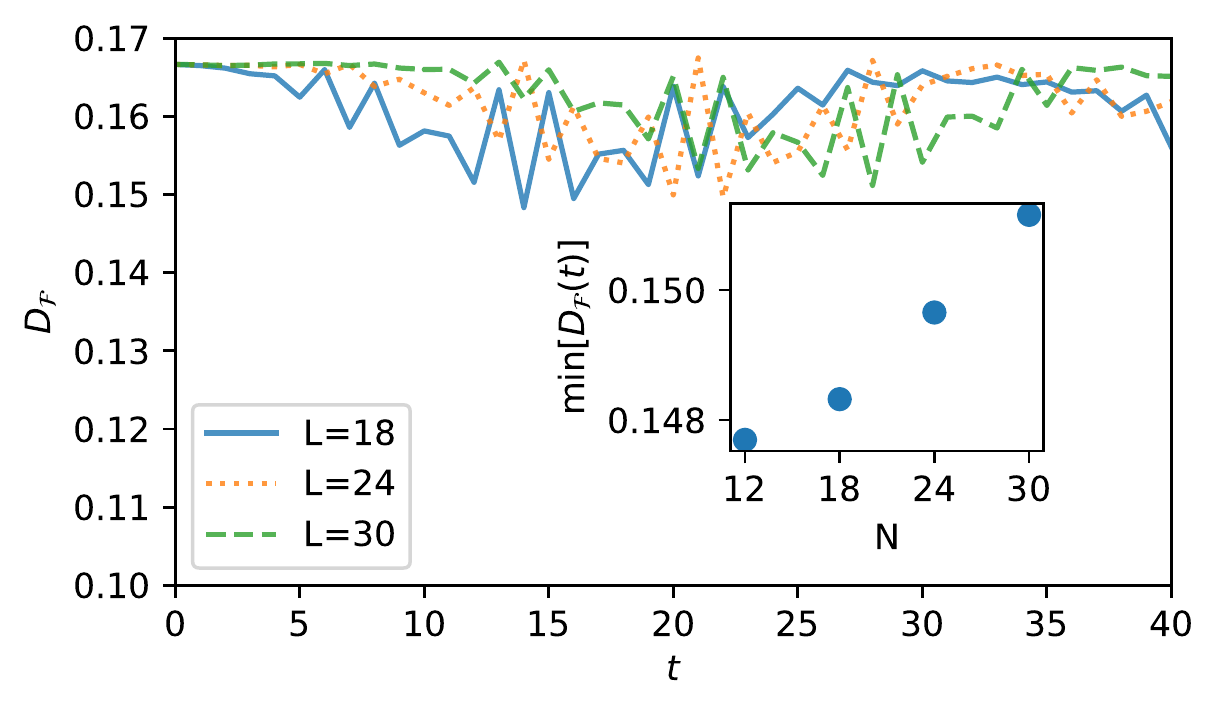}
	\caption{Temporal behavior of interaction distance under quench dynamics from $\zng$ state ($V_i=8$) to the $\zf$ phase ($V_f=-5$) with fixed $U=-15$. Data is for system sizes $N=18,24,30$. Inset shows the scaling of minimum $\df$ as a function of system size.}
	\label{sup-fig:finite-size}
\end{figure}

We find that, in order to yield sensible results from our local Wick’s decomposition in the OBC case, we need to use odd system sizes to properly capture the $\zf$ ground state of the form $\ket{1010...101}$ and with $N$ divisible by three to allow for the $\zng$ phase. This ensures that excitations will be localized at the edges and the bulk behavior will accurately reflect the underlying physics. Wick's decomposition, implemented using formula \eqref{eq:W}, can conveniently probe this. Here, due to OBCs, we evaluate local correlations only within the bulk of the system in order to avoid the boundary effects where the Rydberg blockade is slightly weaker. The Gaussianity phase diagrams, depicted in Fig.~\ref{fig:suppOBC}, demonstrate excellent agreement with the PBC results in terms of both interaction distance and local Wick’s decomposition in Eq.~\eqref{eq:W}.

Finally, we investigated the interaction distance in larger system sizes undergoing quench dynamics from the $\zng$ phase to $\zf$ phase. As can be seen in Fig.~\ref{sup-fig:finite-size}, the non-Gaussianity becomes more pronounced with increasing system size at moderate times $t\lesssim 15$. As the system size increases, the oscillation amplitudes are observed to decrease. Thus, in large system sizes, we anticipate essentially constant $\df$ value with small fluctuations around it.

\bibliography{ref}

\end{document}